# GLOBAL ENTREPRENEURSHIP MONITOR VERSUS PANEL STUDY OF ENTREPRENEURIAL DYNAMICS: COMPARING THEIR INTELLECTUAL STRUCTURES

**Antonio Rafael Ramos-Rodríguez\* & Salustiano Martínez-Fierro & José Aurelio Medina-Garrido & José Ruiz-Navarro**

Department of Management, University of Cádiz, Avda. de la Universidad s/n, 11402 Jerez de la Frontera, Cádiz, Spain

\* Corresponding author: e-mail: rafael.ramos@uca.es

This is the submitted version accepted for publication in the "International Entrepreneurship and Management Journal". The final published version can be found at:

https://doi.org/10.1007/S11365-013-0292-1

We acknowledge that Springer holds the copyright of the final version of this work. Please, cite this paper in this way:

Ramos-Rodríguez, A. R., Martínez-Fierro, S., Medina-Garrido, J. A., & Ruiz-Navarro, J. (2015). Global entrepreneurship monitor versus panel study of entrepreneurial dynamics: comparing their intellectual structures. International Entrepreneurship and Management Journal, 11(3), 571–597. https://doi.org/10.1007/s11365-013-0292-1

**Abstract**

In the past 15 years, two international observatories have been intensively studying entrepreneurship using empirical studies with different methodologies: GEM and PSED. Both projects have generated a considerable volume of scientific production, and their intellectual structures are worth analyzing. The current work is an exploratory study of the knowledge base of the articles generated by each of these two observatories and published in prestigious journals. The value added of this work lies in its novel characterization of the intellectual structure of entrepreneurship according to the academic production of these two initiatives. The results may be of interest to the managers and members of these observatories, as well as to academics, researchers, sponsors and policymakers interested in entrepreneurship.

# GLOBAL ENTREPRENEURSHIP MONITOR VERSUS PANEL STUDY OF ENTREPRENEURIAL DYNAMICS: COMPARING THEIR INTELLECTUAL STRUCTURES

## 1. INTRODUCTION

Since the early 1980s (Veciana, 1999) entrepreneurship has become one of the most popular and active fields of research within the more general field of business administration or management. Thus the number of articles on entrepreneurship appearing in the most important management journals is growing (Busenitz et al., 2003), and many journals have published special editions on the topic. Journals specializing in this field also have a growing academic impact and visibility, as is clear from the increasing number that are included in the ISI databases (e.g., Journal of Business Venturing, Small Business Economics, Journal of Small Business Management and Entrepreneurship: Theory and Practice).

On the other hand, it is generally accepted that firm creation (entrepreneurship) is a young and emerging discipline. Authors describe the research as in a pre-theoretical state (Déry and Toulouse, 1996), or in its adolescence (Grégoire, Déry and Béchard, 2001), or claim it has made only limited progress toward its consolidation as a discipline of knowledge (Aldrich and Baker, 1997; Romano and Ratnatunga, 1996; Busenitz et al., 2003). Nor have authors reached a consensus about its content, theoretical development or constructs (Shane and Venkataraman, 2000; Busenitz et al., 2003; Sarasvathy and Venkataraman, 2011).

Due to its heterogeneous and multi-disciplinary nature intense debate has raged about how difficult it is to identify net contributions to its development, with some authors claiming that entrepreneurship is now a broad label housing "a hodgepodge of research" in need of a theoretical framework that explains and predicts the empirical phenomena (Shane and Venkataraman, 2000).

When a discipline or field of research reaches a certain level of development, academics start to focus on the literature they themselves have generated as an object of research. In particular, when a body of literature develops, it is useful to pause, take stock of the work that has been done and identify new directions and challenges for the future (Low and MacMillan, 1988). The objective at this point is to obtain a general view of the scientific production by looking through the rear-view mirror (White and McCain, 1998).

A bibliometric study can correct researchers' erroneous perceptions about the knowledge base that they are using. For example, researchers may generally agree that a particular author has been or is fundamental to a particular school of thought, but a citation analysis could reveal that this author is cited much less frequently than others. The initial perception may then be discarded, or questioned, and researchers may seek to explain why, if the author is so important, colleagues do not cite his or her work more often (Callon et al., 1993).

Bibliometric studies are helpful for reconstructing the history of a science, questioning traditional dogmas and correcting errors of perception. Moreover, in the sociology of the science, bibliometric studies can shed valuable light on scientific communication. Study of the references may reveal certain problems of isolation or openness toward the outside (from both the thematic and geographic perspectives), the diffusion of new ideas, and the existence of barriers between the applied and fundamental sciences to the specialists and the scientific community to which they belong (Ferreiro, 1993).

Nevertheless, bibliometric methods, and in particular citation analysis, also have their limitations. For example, only documents contained in databases with a citation index structure can be examined, and the way authors use and abuse citations is also problematic. In this research, the authors use the journal article as unit of analysis, so partially overcoming the first limitation. Consequently, the results generated are more reliable since the use of citations is a generally standardized and responsible practice and is subject to the critical analysis of the journal's reviewers.

For this purpose, the current authors carried out an exploratory study of the intellectual structure of the knowledge bases of the scientific production generated by two international projects that have intensively studied entrepreneurship using empirical studies with a global scope.

This work begins with a review of the antecedents, objectives, theoretical models, methodologies and principal results of the projects Global Entrepreneurship Monitor (GEM) and Panel Study of Entrepreneurial Dynamics (PSED). The authors then describe the procedure they use to obtain a representative selection of the research generated by each project, and use bibliometric methods based on co-citation indices to identify the most cited documents, analyze their relations, and subsequently carry out a general and comparative description of their knowledge bases.

The value added of this work lies in its novel characterization of the structure of the knowledge base of entrepreneurship according to the scientific production generated by the two most important international observatories. To date, the literature has not seen studies of a similar nature or scope, so the results of this work may be of interest to the managers and members of the two observatories, as well as academics, researchers, sponsors and policymakers interested in entrepreneurship.

## 2. GLOBAL ENTREPRENEURSHIP MONITOR: ANTECEDENTS, OBJECTIVES, MODEL AND RESULTS

The GEM (Global Entrepreneurship Monitor) project was set up by Babson College (United States) and the London Business School (United Kingdom) in 1999 as an observatory of individuals' attitudes toward entrepreneurship and to evaluate entrepreneurial activity in a large number of countries. The aim was to overcome the lack of international information about entrepreneurial activity (Reynolds et al., 2004a) and to promote research into this topic. In its first year 10 countries participated, and over time the number of participating countries has steadily increased, reaching 69 in the 2012 edition.

The project started to take shape in 1997, and in the following year an initial pilot study ran, with six countries participating. In 2005, the founding institutions and the national GEM teams set up the Global Entrepreneurship Research Association. The increasing importance of this project is clear not only from the growing number of countries participating but also from its annual budget, which rose to US$9 million in 2012, and from the number of surveys it carries out of the general population and experts.

When the project started in 1999 entrepreneurship research suffered from a number of empirical weaknesses (Sternberg and Wennekers, 2005):

- there were no harmonized, internationally comparable data on entrepreneurial activity
- the available statistics on entrepreneurship were not sufficiently up-to-date and did not contain information about the entrepreneurial qualities of the population. This information is crucial for the design of policies to promote new firm creation
- internationally comparable detailed information about the entrepreneurship process was also unavailable.

Figure 1. Theoretical model of GEM project

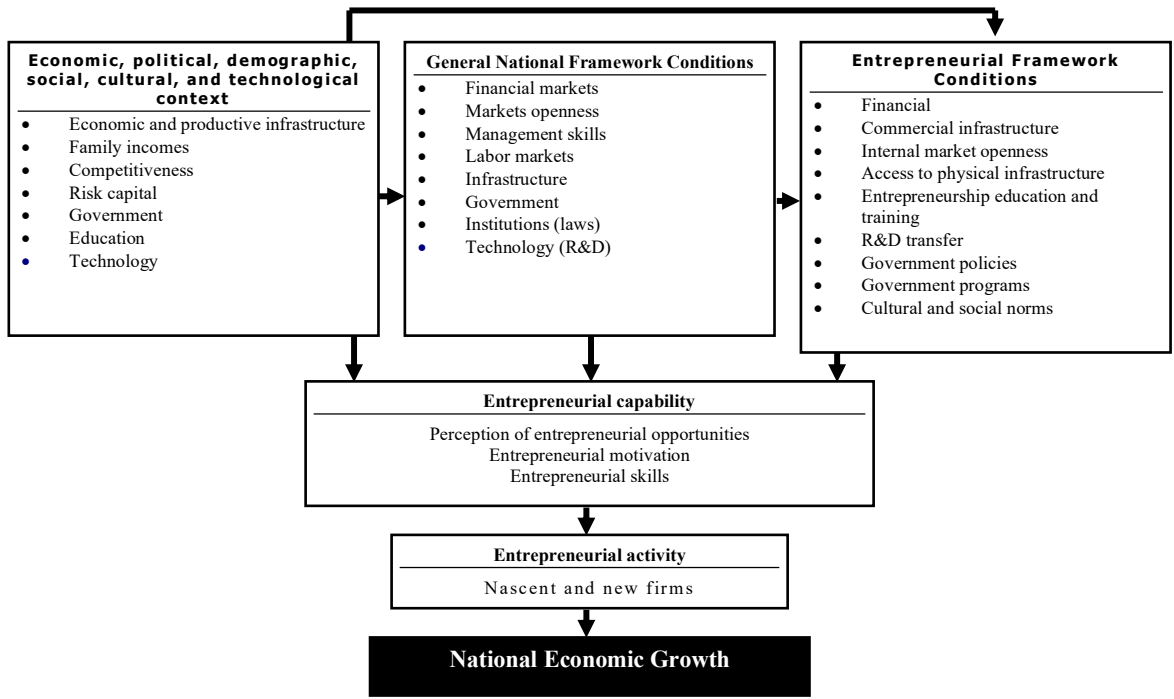

Source: Adapted from Reynolds et al. (2005)

The project aims to analyze the relation between entrepreneurship and the economic growth of countries. Researchers can use its results to compare entrepreneurial activity between countries, estimate the role of entrepreneurship in the economic growth of the country, identify the factors explaining the different levels of entrepreneurial activity between countries, and facilitate efficient and effective policies for promoting entrepreneurship (Reynolds et al., 2005). Thus the ultimate goal of the project is not simply to obtain data but also to analyze them and attempt to answer the following questions (Sternberg and Wennekers, 2005):

- to what extent does the level of entrepreneurial activity vary between countries, and how much does this level change over time?
- why are some countries more entrepreneurial than others?
- what kind of policies can increase the national level of entrepreneurial activity?
- what is the relation between entrepreneurship and economic growth?

The initial model designed to analyze the influence of entrepreneurial activity on a country's economic growth has this latter variable as dependent variable (Figure 1). The starting point is an analysis of the cultural, political and social factors in a particular country that influence the general conditions, the business environment, and the specific conditions impacting on firm creation. The model also assumes that the specific conditions are affected by the general conditions of the business environment. Both the factors and the two groups of conditions influence the capacity to create firms, which is understood in terms of motivation, ability and the perception of business opportunities (Ramos-Rodriguez et al., 2010). The capacity to create firms will give rise to the creation of new ventures, and will consequently have positive repercussions for the country's economic growth.

The GEM project includes the whole process of new firm creation in its model, including all the stages from the firm's conception to its establishment in the market (Figure 2).

Figure 2. The entrepreneurial process according to the GEM model

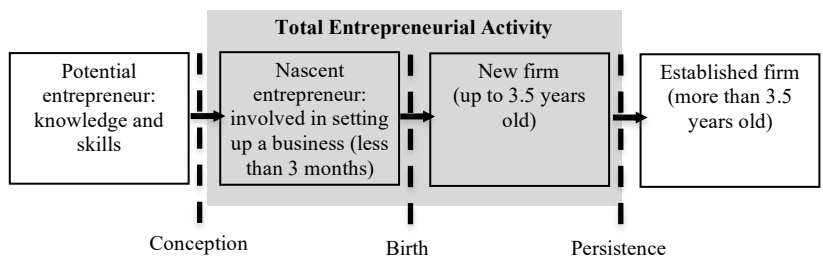

Source: Adapted from Reynolds et al. (2005)

Political, social and economic factors are continually influencing the firm creation process described in the above figure, so the economic situation, entrepreneurial climate and attitudes toward firm creation in a particular country should be considered when analyzing differences between countries and national economic development over time (Reynolds et al., 2005).

GEM uses three sources of information: a survey of the adult population, a panel of experts, and sources of secondary information from each country.

Much of its efforts are concentrated in the survey of the adult population in each participating country. These surveys provide data on the participation of the population in new firm creation, in particular its best-known indicator, the Total Entrepreneurial Activity (TEA) of each country. An entrepreneur according to GEM is any adult who is involved in the process of starting a firm of their property and/or is currently running a young firm, or is the owner of such a firm (Reynolds et al., 2005). These surveys have a minimum of 2,000 respondents in each country each year.

The panel of experts in each country consists of a select group of people familiar with the entrepreneurial phenomenon because of their academic background and experience. A minimum of 36 experts sit on each country's panel. These experts respond to closed questions, but also open questions asking for their opinion about the elements and factors favoring and/or inhibiting firm creation in their country. From the information provided, GEM identifies the conditions influencing firm creation in that country (Entrepreneurial Framework Conditions, EFCs) from a group of nine, previously established, conditions.

The sources of secondary information that GEM tends to use are World Bank (2002), the Global Competitive Reports (Schwab and Sachs, 1997, 1998), and World Economic Forum (2002), as well as others specific to each country.

A global team and the coordinators of each participating country coordinate the procedure for collecting the information at the global level and in each of the countries. They guide the national teams and the regional teams in those countries where the project also runs on a regional level.

The original model presented above has been evaluated and revised over time, and a number of modifications have been made to improve understanding of the influence of new firm creation on a country's economic development (Bosma et al., 2009; Bosma and Levie, 2010). In this respect, and considering that the potential contribution of entrepreneurs to a country's economic development depends on its stage of development (Wennekers et al., 2005; Gries and Naude, 2008) and that the Global Competitiveness Index has evolved considerably since the end of the 1990s (Bosma et al., 2012), a more subtle distinction between the stages of economic development has been introduced following Porter's typology, which distinguishes between factor-driven, efficiency-driven and innovation-driven countries (Porter et al., 2002).

The revised model also incorporates the three components of the entrepreneurship profile: entrepreneurial attitudes, entrepreneurial activity, and entrepreneurial aspirations. These are seen as a "black box" that contributes to innovation and employment generation in an economy (Bosma et al., 2012). The novelty with respect to the original model here is the inclusion of aspirations alongside the entrepreneurial attitudes and activity (Figure 3).

Figure 3. Revised GEM model

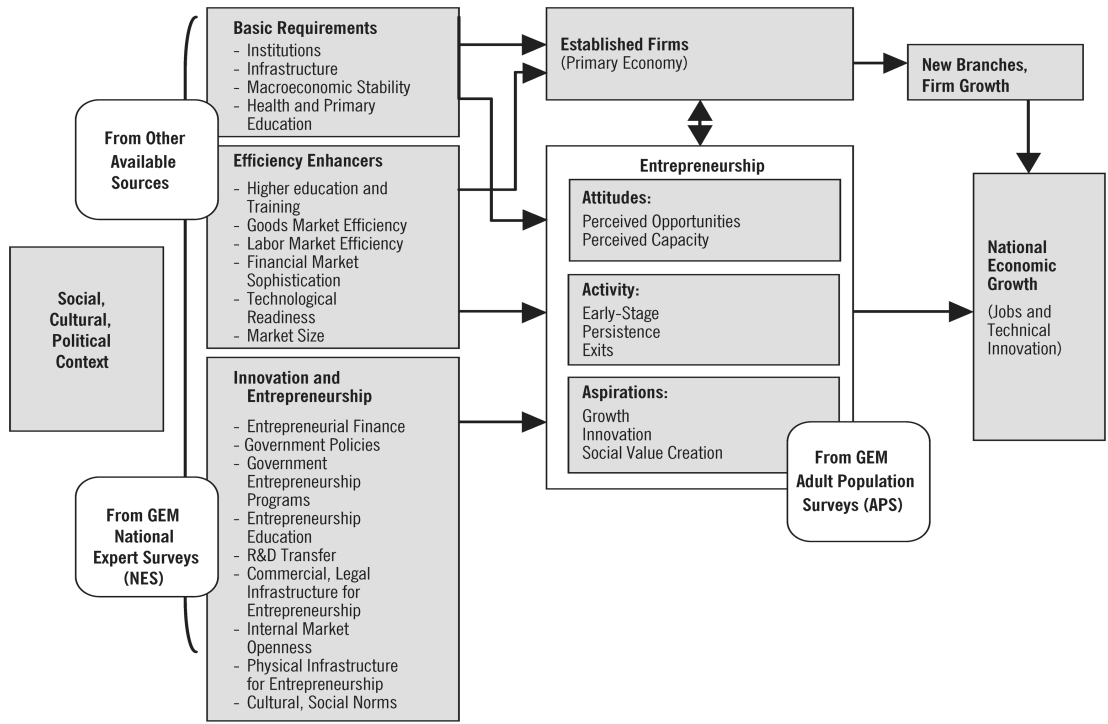

Source: Bosma and Levie (2010)

With the GEM project up and running and the data available to researchers, the Global GEM reports and the national GEM reports for each of the participating countries have been published annually. Some of the countries also have regional teams, which follow the same procedure for a specific region and publish a corresponding annual report. To complement all this information, teams have also published reports focusing on specific topics such as women and entrepreneurship, high-growth entrepreneurship, new business financing, social entrepreneurship, and so on.

With regard to academic and scientific research, a growing number of studies using the GEM databases are appearing in international journals. Because of this expanding research there is a need to review the work published to date to gain a greater understanding of the utility of the project. Thus Urbano et al. (2010) analyze the most important academic work that uses GEM databases in the period 1999-2006. They find 50 studies, including 15 journal articles (of which 11 are in Journal Citation Report impact journals), with the rest being work presented in conferences. More recently, Álvarez and Urbano (2011) analyze the content and evolution of articles that use GEM databases in entrepreneurship journals included in the JCR, finding 48 articles.

## 3. PANEL STUDY OF ENTREPRENEURIAL DYNAMICS: ANTECEDENTS, OBJECTIVES, MODEL AND RESULTS

The PSED project has its roots in a 1993 study of adults in Wisconsin that had financial support from the Wisconsin Housing and Economic Development Authority. The fundamental design of this study involved selecting representative samples of adults to locate people involved in new firm creation, obtain detailed information about their efforts, and carry out follow-up interviews to determine the outcomes. The initiative proved successful, and the Survey Research Center at the University of Michigan used the same methodology and expanded the project to a representative sample of households across the United States in 1996. The research design confirmed its reliability, and this led to the creation of the Entrepreneurial Research Consortium, which brought together more than 120 academics from 34 research centers studying entrepreneurship. The Consortium provided financial backing to run the first Panel Study of Entrepreneurial Dynamics (PSED I) in 1998 (Curtin and Reynolds, 2007).

The PSED research project is a powerful resource for advancing academic understanding of the firm creation process, and focuses on the development of the new businesses during their infancy (Reynolds and Curtin, 2007). It was designed to provide reliable and systematic longitudinal data about the early years of the entrepreneurial process. What distinguishes PSED from other studies is that this project identifies individuals involved in firm creation and obtains generalizable information from them about the characteristics of the process, the population involved in firm creation, and the activities and other attributes of the nascent firms. Its model incorporates theoretical knowledge about the demographic and cognitive characteristics of the entrepreneurs, opportunity recognition, the formation of teams, networks, the resources that form part of the entrepreneurial process, and the characteristics of the environment that conditions the process (Gartner et al., 2004).

To date the project has completed two editions: 1998 and 2005. In each edition the surveys are longitudinal, with the same individuals repeating three times in the first edition and twice in the second (Reynolds, 2000; Gartner et al., 2004). Almost 50 countries participated in PSED during this period. In this model, researchers analyze the individuals or teams involved in the start-up process, what happens during that process, the typical characteristics of the individuals, teams and activities, and whether the start-ups survive or fail (Reynolds and Curtin, 2007).

The first edition of the project ran from 1998 to 2000. It consisted of a first detailed interview of adults who had started new firms and a further three follow-up interviews in a three-year period (Reynolds, 2000; Gartner et al., 2004). The first part of this process was carried out by the University of Wisconsin and obtained financial support in the shape of two grants from the National Science Foundation for two samples, one of women and the other of minorities. After the initial survey and the first follow-up survey, the research laboratory at the University of Wisconsin closed, and thanks to financial support from the Ewing Marion Kauffman Foundation, the University of Michigan's Institute for Social Research assumed responsibility for collecting the data and organized the second and third follow-up interviews.

Other longitudinal studies based on the PSED I paradigm soon followed: in Canada (Diochon et al., 2007; Menzies et al., 2002), the Netherlands (Wolters, 2000), Norway (Alsos and Kolvereid, 1998), and Switzerland (Delmar and Davidsson, 2000).

The research design of PSED II improved on the procedures of the previous version. The project obtained funding from the Ewing Marion Kauffman Foundation, complemented by funds from the US Small Business Administration in 2004, to carry out a second edition of the project. The University of Michigan's Institute for Social Research again collected the data.

There were a number of reasons for running a second panel (Reynolds and Curtin, 2007). First, PSED I ran in the period 1998-2000, at the height of the dot-com firms, making it advisable to analyze firms in a more normal period. Second, a series of methodological improvements in the light of the considerable experience of the PSED I and GEM projects, led to a much improved and more efficient research protocol, which provided more reliable measures and improved measures for some critical topics. Third, researchers had found that ethnic groups differed in terms of their participation and experiences, and this required closer attention. And finally, the large number of factors that affected the results suggested that a larger sample of nascent firms would be very valuable for a multivariate analysis.

In both PSED editions researchers used a longitudinal methodology with follow-up interviews during a certain period of time. There were three follow-up interviews in the first edition and two in the second. Below the authors describe the process followed in PSED II, since this is the most up-to-date and improved version of the project.

During the autumn of 2005 and early winter of 2006, 31,845 people were interviewed, with 1,214 identifying themselves as nascent entrepreneurs. The second stage ran 12 months later, and consisted of a telephone interview lasting 60 minutes in which the respondent provided information about a wide range of topics concerning the birth of their new firm (Reynolds and Curtin, 2009): type of business, initiation of activities in the name of the new firm, inclusion in business registers, nature of entrepreneurial team and collaboration in networks, sources and level of financial support, specific environment, competitive strategy, expectations of growth, motivations, perspectives, self-descriptions, and antecedents and family context of nascent entrepreneur.

The third stage consisted of another 60-minute telephone interview two years after the first interview. The topics discussed depended on the firm's status at that particular time. Thus entrepreneurs who had given up on their venture answered only a few questions about the start of the activity and their reasons for giving up, while the rest received the whole interview and had the opportunity to update their records with new information about the firm's operations. Entrepreneurs who stated that they were running a new firm in the previous stage were asked some additional modules of questions concerning the cost structure to provide data to estimate the labor productivity. These modules as well of those concerning the firm's organizational structure, were designed to facilitate comparison with similar modules from the panel study of new firms sponsored by Kauffman (Haltiwanger et al., 2007; Mathematica Policy Research, 2007).

The data generated by the two editions of the PSED project have given rise to the publication of the results of the project itself, by Reynolds and Curtin, along with a large number of articles and books. These authors' report from 2007 lists 7 books, 8 book chapters, 45 articles in peer-reviewed journals, 6 research reports and 63 professional presentations up to that time.

Over time the work using PSED panel data has continued to grow, and according to the latest data available on the project's website (from 2012), the output has reached 96 articles in peer-reviewed journals, 12 books, 66 book chapters, 18 doctoral theses, 10 research reports, and a large number of presentations in national and international congresses.

## 4. METHODOLOGY

### 4.1 Data

The first step in carrying out this research involved identifying a body of research representative of the scientific production generated around the GEM and PSED projects. The authors considered journal articles and research notes as sources of scientific documentation for this analysis, rather than books, doctoral theses and congress acts, because they consider the former "certified knowledge". This is the term commonly used to describe knowledge that has been subject to the critical review of colleagues and has been accepted for publication (Callon et al., 1993). Moreover, as the authors carried out a citation analysis later, the use of citations in journal articles and research notes is standard practice, which increases the reliability of the results.

Given the nature of the techniques used in this research, the authors limited their search for documents to databases with a citation index structure, in other words, databases that codify the Bibliography or References section in each article or research note. Consequently, in order to obtain the two collections of research articles to represent the two observatories of entrepreneurship analyzed, the authors first carried out a bibliographic search in the Social Science Citation Index (SSCI) and extracted all articles published up to January 2013 (period of data collection) that contained in the title, abstract or keywords, any of the following terms:

- for GEM: ("global entrepreneurship monitor" OR "gem data" OR "gem project" OR "gem study") AND (entrepreneur*)
- for PSED: ("panel study of entrepreneurial dynamics" OR psed) AND (entrepreneur*)

The authors took all the articles collected with this criterion and built a file (henceforth, the "citing sample"), and then extracted another file from this file with all the bibliographic references cited by those articles (henceforth, the "cited sample").

There were a large number of inconsistencies in the coding of the citations in this database, and the bibliometric software used[1] to count the citations considers them as simple chains of characters, so the authors had to refine the citations manually, paying particular attention to the spelling of authors' names, the titles of the journals, and books with several editions, to guarantee the accuracy of subsequent counts[2].

## 4.2 Bibliometric techniques: citation analysis and co-citation analysis

The bibliometric techniques used in this work are citation analysis and co-citation analysis. Citation analysis is based on the premise that authors cite documents that they have used to carry out their research and that they consider important for their work. Thus it is reasonable to assume that when analyzing the literature generated by a large number of authors working within a particular scientific discipline, the more frequently a work is cited the greater its influence on the construction and development of that field. Applying this technique leads to a frequency distribution that ranks the documents in function of their citation frequency. Researchers tend to use this frequency as an indicator of a work's influence, visibility or impact, but should never equate this visibility or impact with quality.

---

[1] The authors used BIBEXCEL, an application for handling bibliographic data developed by Prof. Olle Pearson at the University of Umeå, Sweden. It can be downloaded free of charge at: www.umu.se/inforsk/Bibexcel/.
[2] For example, Barney (1991) is coded variously as BARNEY J, 1991, V17, P99, J MANAGE and as BARNEY JB, 1991, V17, P99, J MANAGE, so the current authors had to standardise this author's initials by hand. Inconsistencies in the coding are also frequent in the abbreviations of journal names.

Document co-citation analysis, for its part, starts from a calculation of the number of times two particular documents are cited together (i.e., in the same work) by later research. This co-citation frequency can be interpreted as a measure of how similar the two works are in terms of content. This measure of proximity is useful for identifying groups of authors, research problems and methodologies, and can shed light on how these groups inter-relate (Pilkington and Liston-Heyes, 1999). In the case of author co-citation analysis, White and Griffith (1981), McCain (1990) and White and McCain (1998) describe this procedure in detail and a large number of studies confirm its validity for exploring the intellectual structure of a scientific discipline (Small, 1973; White and Griffith, 1981; McCain, 1986; Culnan, O'Reilly and Chatman, 1990; White and McCain, 1998; Ying, Gobinda and Schubert, 1999; Reader and Watkins, 2001).

Co-citation analysis involves counting the frequency with which any pair of studies from among the most influential studies identified in the previous stage are cited together (Figure 4). These frequencies go into a square, symmetric matrix (raw co-citation matrix), the dimension of which is equal to the number of cited documents selected and the main diagonal of which is undefined[3].

---

[3] Calculating the number of articles listing the same document twice in their Reference section makes no sense (it is always zero).

Figure 4. Obtaining the co-citation frequencies

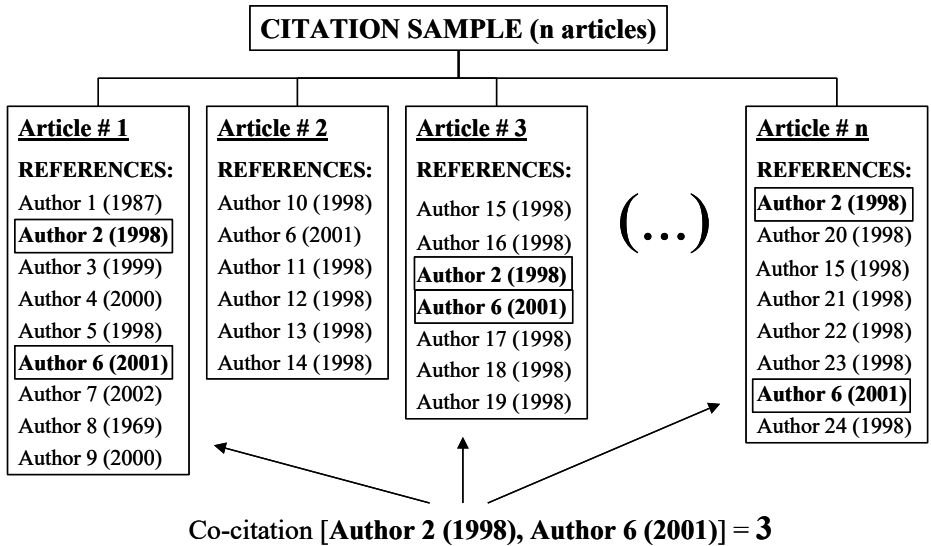

Source: The Authors

Researchers usually have to deal with a very large number of cited works (3,347 different documents in GEM, and 1,515 in PSED), so the procedure described above requires a cut-off point or citation threshold, which means that the researchers only analyze references that obtain more than a certain number of citations. In this work the authors consider articles receiving five citations or more for GEM and three citations or more for PSED. This makes their citation thresholds similar, at 19.95% and 21.12%, respectively. Consequently, of the 3,347 documents cited by the articles published in the framework of the GEM project the authors analyzed 118 documents, which were referenced 979 times in total (19.95%). In the case of PSED, the authors analyzed 92 studies, which were referenced 417 times (21.12%).

Setting the citation thresholds is a difficult design problem because it influences the final results. To date, no studies offer reliable methodological guidance on this point, so the authors took their decision after carrying out a large number of tests and by a process of trial and error until they obtained co-citation matrices of a suitable size for handling by computer and subsequent interpretation.

Finally, the authors used tools from social network analysis (e.g., Netdraw) to produce the graphs of relations based on the co-citation frequency.

To handle the data the authors used Bibexcel, a specialist bibliometric software developed by Olle Persson at the University of Umeå, in Sweden.

## 5. RESULTS AND DISCUSSION

After applying the search criteria described in the previous section the authors identified, in the SSCI, 86 articles linked to GEM and 34 to PSED. The first group of articles made a total of 4,908 bibliographic references to a total of 3,347 different documents, while the PSED group made a total of 1,974 references to 1,515 different documents.

Table 1 lists the 118 documents (citation threshold of 19.95%) most cited by the 86 articles of the citing sample in the GEM project.

Table 1. Intellectual base of GEM (most cited documents)

| Citations received | Documents |
|---|---|
| 46 | Reynolds P, 2005, V24, P205, Small Bus Econ |
| 26 | Wennekers S, 2005, V24, P293, Small Bus Econ |
| 19 | Davidsson P, 2003, V18, P301, J Bus Venturing |
| 18 | Arenius P, 2005, V24, P233, Small Bus Econ |
| 18 | Wennekers S, 1999, V13, P27, Small Bus Econ |
| 18 | Shane S, 2000, V25, P217, Acad Manage Rev |
| 17 | North D, 1990, I I Change Ec Perfor |
| 17 | Reynolds P, 2002, Global Entrepreneurs |
| 15 | Carree M, 2002, V19, P271, Small Bus Econ |
| 15 | Sternberg R, 2005, V24, P193, Small Bus Econ |
| 14 | Baumol W, 1990, V98, P893, J Polit Econ |
| 14 | Schumpeter J, 1934, Theory Ec Dev |
| 13 | Bosma N, 2008, Global Entrepreneurs |
| 13 | Acs Z, 2005, Global Entrepreneurs |
| 13 | Wong P, 2005, V24, P335, Small Bus Econ |
| 12 | Minniti M, 2006, Global Entrepreneurs |
| 12 | Krueger N, 2000, V15, P411, J Bus Venturing |
| 11 | Van Stel A, 2007, V28, P171, Small Bus Econ |
| 11 | Ardichvili A, 2003, V18, P105, J Bus Venturing |
| 11 | Ajzen I, 1991, V50, P179, Organ Behav Hum Dec |
| 10 | Storey D, 1994, Understanding Small |
| 10 | Delmar F, 2000, V12, P1, Entrep Region Dev |
| 10 | Levesque M, 2006, V21, P177, J Bus Venturing |
| 10 | Arenius P, 2005, V24, P249, Small Bus Econ |
| 10 | Parker S, 2004, Ec Self Employment E |
| 10 | Van Stel A, 2005, V24, P311, Small Bus Econ |
| 10 | Baumol William J, 2002, Free Market Innovati |
| 10 | Reynolds P, 1994, V28, P443, Reg Stud |
| 10 | Bygrave W, 2003, V5, P101, Venture Capital Int |
| 9 | Kirzner I, 1973, Competition Entrepre |
| 9 | Verheul I, 2006, V18, P151, Entrep Region Dev |
| 9 | Shapero A, 1982, P72, Ency Entrepreneurshi |
| 9 | Hofstede G, 2001, Cultures Consequence |
| 9 | Reynolds P, 2001, Global Entrepreneurs |
| 9 | Reynolds P, 2003, Global Entrepreneurs |
| 9 | Bosma N, 2009, Global Entrepreneurs |
| 9 | Minniti M, 2005, Global Entrepreneurs |
| 9 | Reynolds P, 1999, Global Entrepreneurs |
| 9 | Audretsch D, 2001, V10, P267, Ind Corp Change |
| 8 | Mcclelland D, 1961, Achieving Soc |
| 8 | Blau D, 1987, V95, P445, J Polit Econ |
| 8 | Acs Z, 2008, V31, P305, Small Bus Econ |
| 8 | North D, 2005, P1, Princ Econ Hist W Wo |
| 8 | Busenitz L, 2000, V43, P994, Acad Manage J |
| 8 | Klapper L, 2006, V82, P591, J Financ Econ |
| 8 | North D, 1994, V84, P359, Am Econ Rev |
| 8 | Grilo I, 2006, V26, P305, Small Bus Econ |
| 8 | Bosma N, 2010, Global Entrepreneurs |
| 7 | Uhlaner L, 2007, V17, P161, J Evol Econ |
| 7 | Audretsch D, 2002, Entrepreneurship Det |
| 7 | Robinson P, 1994, V9, P141, J Bus Venturing |
| 7 | Carree M, 2007, V19, P281, Entrep Region Dev |
| 7 | Levie J, 2008, V31, P235, Small Bus Econ |
| 7 | Cooper A, 1987, V11, P11, Am J Small Business |
| 7 | Hayton J, 2002, V26, P33, Entrep Theory Pract |
| 7 | Porter M, 2002, P16, Global Competitivene |
| 7 | Davidsson P, 2006, V2, Foundations And Trends In Entrepreneurship |
| 7 | Hofstede G, 1980, Cultures Consequence |
| 7 | Grilo I, 2005, V1, P441, Int Entrepreneurship |
| 7 | Shane S, 2000, V11, P448, Organ Sci |
| 6 | Etzioni A, 1987, V8, P175, J Econ Behav Organ |
| 6 | Evans D, 1989, V79, P519, Am Econ Rev |
| 6 | Audretsch D, 2002, V36, P113, Reg Stud |
| 6 | Armington C, 2002, V36, P33, Reg Stud |
| 6 | Amoros J, 2008, V4, P381, Int Entrepreneurship |
| 6 | Kirzner I, 1997, V35, P60, J Econ Lit |
| 6 | Aidis R, 2008, V23, P656, J Bus Venturing |
| 6 | Kolvereid L, 1996, V21, P47, Entrep Theory Pract |
| 6 | Reynolds P, 2004, Global Entrepreneurs |
| 6 | Blanchflower D, 2001, V45, P680, Eur Econ Rev |
| 6 | Carree M, 2003, P437, Hdb Entrepreneurship |
| 6 | Djankov S, 2002, V117, P1, Q J Econ |
| 6 | Schumpeter J, 1942, Capitalism Socialism |
| 6 | Acs Z, 2007, V28, P109, Small Bus Econ |
| 6 | Acs Z, 2005, V24, P323, Small Bus Econ |
| 6 | Noorderhaven N, 2004, V28, P447, Entrep Theory Pract |
| 6 | Davidsson P, 1995, V7, P41, Entrep Region Dev |
| 6 | Davidsson P, 1991, V6, P405, J Bus Venturing |
| 6 | Casson M, 1982, Entrepreneur Ec Theo |
| 6 | Gartner W, 1985, V10, P696, Acad Manage Rev |
| 5 | Scherer R, 1991, V3, P195, Entrep Region Dev |
| 5 | Mueller S, 2000, V16, P51, J Business Venturing |
| 5 | Mueller S, 2001, V16, P51, J Bus Venturing |
| 5 | Wennekers S, 2007, V17, P133, J Evol Econ |
| 5 | Chen C, 1998, V13, P295, J Bus Venturing |

| 5 | Koellinger P, 2008, V31, P21, Small Bus Econ |
| --- | --- |
| 5 | Shane S, 2003, Gen Theory Entrepren |
| 5 | Hair J, 1998, Multivariate Data An |
| 5 | Knight F, 1921, Risk Uncertainty Pro |
| 5 | Reynolds P, 1997, V9, P449, Small Bus Econ |
| 5 | Harper D, 2003, Fdn Entrepreneurship |
| 5 | Porter M, 1990, Competitive Advantag |
| 5 | Davidsson P, 1997, V18, P179, J Econ Psychol |
| 5 | Audretsch D, 2004, V38, P949, Reg Stud |
| 5 | Lucas R, 1978, V9, P508, Bell J Econ |
| 5 | Audretsch D, 2000, V10, P17, J Evol Econ |
| 5 | Johnson S, 2002, V92, P1335, Am Econ Rev |
| 5 | Verheul I, 2002, P11, Entrepreneurship Det |
| 5 | Lumpkin G, 1996, V21, P135, Acad Manage Rev |
| 5 | Rocha H, 2005, V24, P267, Small Bus Econ |
| 5 | Wagner J, 2004, V38, P219, Ann Regional Sci |
| 5 | Freytag A, 2007, V17, P117, J Evol Econ |

| 5 | De Carolis D, 2006, V30, P41, Entrep Theory Pract |
| --- | --- |
| 5 | De Clercq D, 2006, V24, P339, Int Small Bus J |
| 5 | Gimeno J, 1997, V42, P750, Admin Sci Quart |
| 5 | Kirzner I, 1979, Perception Opportuni |
| 5 | Tiessen J, 1997, V12, P367, J Bus Venturing |
| 5 | Desoto H, 1989, Other Path Invisible |
| 5 | Becker G, 1964, Human Capital |
| 5 | Blanchflower D, 2000, V7, P471, Labour Econ |
| 5 | Bergmann H, 2007, V28, P205, Small Bus Econ |
| 5 | Acs Z, 2006, Geography, And American Economic Growth, P1, Entrepreneurship |
| 5 | Krueger N, 2000, V24, P5, Entrep Theory Pract |
| 5 | Krueger N, 1994, V19, P91, Entrep Theory Pract |
| 5 | Acs Z, 2007, V28, P123, Small Bus Econ |
| 5 | Evans D, 1989, V97, P808, J Polit Econ |
| 5 | Levie J, 2007, V28, P143, Small Bus Econ |
| 5 | Feldman M, 2001, V10, P861, Ind Corp Change |

Table 2 lists the 92 documents (citation threshold of 21.12%) most cited by the 34 articles of the citing sample in the PSED project.

Table 2. Intellectual base of PSED (most cited documents)

| Citations received | Documents |
| --- | --- |
| 18 | Gartner W, 2004, Hdb Entrepreneurial |
| 11 | Carter N, 2003, V18, P13, J Bus Venturing |
| 11 | Delmar F, 2003, V24, P1165, Strategic Manage J |
| 10 | Shane S, 2000, V25, P217, Acad Manage Rev |
| 10 | Reynolds P, 1997, Entrepreneurial Proc |
| 8 | Cooper A, 1994, V9, P371, J Bus Venturing |
| 8 | Delmar F, 2004, V19, P385, J Bus Venturing |
| 8 | Davidsson P, 2003, V18, P301, J Bus Venturing |
| 8 | Katz J, 1988, V13, P429, Acad Manage Rev |
| 7 | Aldrich H, 1999, Org Evolving |
| 7 | Carter N, 1996, V11, P151, J Bus Venturing |
| 7 | Reynolds P, 2000, V4, P153, Adv Entrepreneurship |
| 6 | Gatewood E, 1995, V10, P371, J Bus Venturing |
| 6 | Vroom V, 1964, Work Motivation |
| 6 | Baker T, 2005, V50, P329, Admin Sci Quart |
| 6 | Gartner W, 1985, V10, P696, Acad Manage Rev |
| 6 | Evans D, 1989, V79, P519, Am Econ Rev |
| 6 | Parker S, 2006, V27, P81, Small Bus Econ |
| 6 | Shaver K, 1991, V16, P23, Entrep Theory Pract |
| 6 | Reynolds P, 2004, V23, P263, Small Bus Econ |
| 6 | Kirchhoff Bruce A, 1994, Entrepreneurship Dyn |
| 5 | Kim P, 2006, V27, P5, Small Bus Econ |
| 5 | Aldrich H, 2001, V25, P41, Entrep Theory Pract |
| 5 | Krueger N, 2000, V15, P411, J Bus Venturing |
| 5 | Knight F, 1921, Risk Uncertainty Pro |
| 5 | Reynolds P, 1992, V7, P405, J Bus Venturing |
| 5 | Gartner W, 2004, P285, Hdb Entrepreneurship |
| 5 | Bruderl J, 1992, V57, P227, Am Sociol Rev |
| 5 | Gimeno J, 1997, V42, P750, Admin Sci Quart |
| 4 | Reynolds P, 2007, V3, Foundations And Trends In Entrepreneurship |
| 4 | Blanchflower D, 1998, V16, P26, J Labor Econ |
| 4 | Delmar F, 2003, V18, P189, J Bus Venturing |
| 4 | Bosma N, 2004, V23, P227, Small Bus Econ |
| 4 | Suchman M, 1995, V20, P571, Acad Manage Rev |
| 4 | Sarasvathy S, 2001, V26, P243, Acad Manage Rev |
| 4 | Kirzner I, 1997, V35, P60, J Econ Lit |
| 4 | Schumpeter J, 1934, Theory Ec Dev |
| 4 | Shane S, 2000, V11, P448, Organ Sci |
| 4 | Reynolds P, 2009, V33, P151, Small Bus Econ |
| 4 | Shane S, 2004, V19, P767, J Bus Venturing |
| 4 | Curtain R, 2004, P477, Hdb Entrepreneurial |
| 4 | Dunn T, 2000, V18, P282, J Labor Econ |
| 4 | Liao J, 2006, V27, P23, Small Bus Econ |
| 4 | Gartner W, 2003, P195, Hdb Entrepreneurship |
| 4 | Brush C, 2008, V23, P547, J Bus Venturing |
| 4 | Bates T, 1990, V72, P551, Rev Econ Stat |
| 4 | Bruderl J, 1998, V10, P213, Small Bus Econ |
| 4 | Gatewood E, 2002, V27, P187, Entrep Theory Pract |
| 4 | Stinchcombe A, 1965, P142, Hdb Org |
| 4 | Baron R, 1998, V13, P275, J Bus Venturing |
| 3 | Bhide A, 2000, Origin Evolution New |
| 3 | Chen C, 1998, V13, P295, J Bus Venturing |
| 3 | Starr J, 1990, V11, P79, Strategic Manage J |
| 3 | Busenitz L, 1997, V12, P9, J Bus Venturing |
| 3 | Shapero A, 1982, P72, Ency Entrepreneurshi |
| 3 | Buttner E, 1997, V35, P34, J Small Bus Manage |
| 3 | Cassar G, 2007, V19, P89, Entrep Region Dev |
| 3 | Shane S, 2003, V13, P257, Human Resource Manag |
| 3 | Birley S, 1994, V9, P7, J Bus Venturing |
| 3 | Bates T, 2005, V20, P343, J Bus Venturing |
| 3 | Meyer J, 1977, V83, P340, Am J Sociol |
| 3 | Hair J, 1995, Multivariate Data An |
| 3 | Manolova T, 2007, V31, P407, Entrep Theory Pract |
| 3 | Gartner W, 2003, P103, New Movements In Entrepreneurship |
| 3 | Gatewood E, 1993, V17, P91, Entrep Theory Pract |
| 3 | Gatewood E, 2004, P153, Hdb Entrepreneurial |
| 3 | Hayek F, 1945, V35, P519, Am Econ Rev |
| 3 | Honig B, 2004, V30, P29, J Manage |
| 3 | Hurst E, 2004, V112, P319, J Polit Econ |
| 3 | Larson A, 1993, V17, P5, Entrep Theory Pract |
| 3 | Low M, 1997, V12, P435, J Bus Venturing |
| 3 | Lichtenstein B, 2007, V22, P236, J Bus Venturing |
| 3 | Holtzeakin D, 1994, V102, P53, J Polit Econ |
| 3 | Evans D, 1989, V97, P808, J Polit Econ |
| 3 | Rotefoss B, 2005, V17, P109, Entrep Region Dev |
| 3 | Robinson P, 1994, V9, P141, J Bus Venturing |
| 3 | Davidsson P, 2006, V2, Foundations And Trends In Entrepreneurship |
| 3 | Cliff J, 1998, V13, P523, J Bus Venturing |
| 3 | Ruef M, 2010, P1, Kauff Found Ser |
| 3 | Curtin R, 2004, Hdb Entrepreneurial |
| 3 | Reynolds P, 2004, P495, Hdb Entrepreneurial |
| 3 | Reynolds P, 2001, Global Entrepreneurs |
| 3 | Douglas E, 2000, V15, P231, J Bus Venturing |

| 3 | Parker S, 2004, Ec Self Employment E |
| 3 | Reynolds P, 2004, P453, Hdb Entrepreneurial |
| 3 | Delmar F, 2000, V12, P1, Entrep Region Dev |
| 3 | Delmar F, 2006, V4, Strategic Organization |
| 3 | Ardichvili A, 2003, V18, P105, J Bus Venturing |

| 3 | Ajzen I, 1991, V50, P179, Organ Behav Hum Dec |
| 3 | Xu H, 2004, V2, P331, Strateg Organ |
| 3 | Wiklund J, 2003, V40, P1919, J Manage Stud |
| 3 | Aldrich H, 1987, P154, Frontiers Entreprene |

## 5.1 Intellectual structure of GEM corpus

Figure 5 shows the structure of the intellectual base of the works most cited in the GEM research. A line linking two points representing studies in the intellectual base means that the two works are co-cited at least 8 times in the citing sample. A co-citation frequency is chosen so that the graph is visually interpretable by the researcher. In this case, the authors decided that the graph was interpretable when it retained studies co-cited at least 8 times in the citing sample.

Figure 5. Structure of intellectual base of SSCI production in GEM project (raw co-citation≥8)

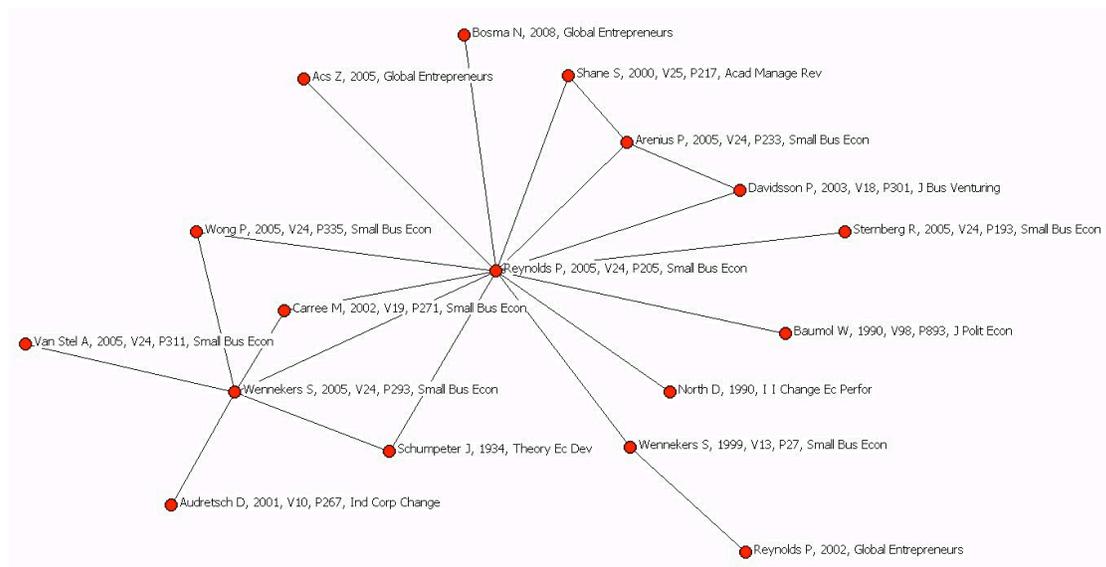

As is clear from the figure, Reynolds et al. (2005) lies right at the heart of the network of co-citations. This work is the most cited work in this field, and is the first work to describe the GEM project and particularly its methodology. This is why Reynolds et al. (2005) became the cornerstone of this field, and practically an automatic reference for any research using data from the GEM project.

Another influential work is Wennekers et al. (2005), which has the above-mentioned Prof Reynolds as co-author. This work uses GEM project data to demonstrate the U-shaped relation between a country's level of entrepreneurial activity and its level of economic development. The study is founded on the line of research begun by Wennekers and Thurik (1999) on entrepreneurship and economic growth, which in turn was based on various fields: historical view of entrepreneurship, theory of macro-economic growth, theory of growth, industrial economy, evolutionary economy, history of economic growth, and management in large organizations. Oddly, Wennekers and Thurik (1999) is less cited than the later Wennekers et al. (2005), but it is co-cited with Reynolds et al. (2005) more frequently.

Analyzing the works co-cited with Reynolds et al. (2005), two works stand out because of their interconnection with this work, that is, the number of times they are co-cited with it: Davidsson and Honig (2003) and Arenius and Minniti (2005). The former analyses how social and human capital influence new firm creation, while the latter analyses how certain perceptual variables correlate with an individual's decision to start a new business. These perceptual variables are: to be alert to opportunities, fear of failure, and confidence in one's entrepreneurial abilities. The first of these perceptual variables, to be alert to opportunities, is precisely what connects this work to Shane and Venkataraman (2000). These latter authors stress the importance of entrepreneurship as a field of research and analyze conceptually the existence, discovery and decision to exploit business opportunities.

Other important studies with a large number of citations are: North (1990), which uses the theory of institutional change to explain how an institution's past behaviors influence its present and future behaviors and ultimately its economic performance; Baumol (1990), who expands the Schumpeterian concept of innovation to include non-productive activities, including those that are destructive to the economy, that entrepreneurs may take up motivated by the right reward system; Sternberg and Wennekers (2005), who describe the theoretical and methodological bases of the GEM project and stress that (1) the impact of entrepreneurial activity differs depending on the country's level of economic development, (2) the creation of high-growth firms boosts the spread of knowledge and economic growth, and (3) entrepreneurship can only be studied by considering the contextual conditions at the regional level; Carree et al. (2002), who analyze the relation between the ownership of the business and economic development; Schumpeter (1934), which is a classic and practically automatic reference, since it is the first work to consider the entrepreneur as an innovator and to analyze the relation between cycles of economic development and the existence of innovative entrepreneurs; and Wong et al. (2005), who uses GEM data to study firm creation and technological innovation as determinants of economic growth. The last three of these studies have a high co-citation frequency with both Reynolds et al. (2005) and Wennekers et al. (2005).

Two works are intriguing: Van Stel et al. (2005) and Audretsch and Thurik (2001). These two studies have a relatively low number of citations (10 and 9 respectively), yet they remain in the graph despite the co-citation threshold of 8. Other studies with a higher citation frequency have disappeared from the graph, but these studies remain because of their high co-citation frequency with Wennekers et al. (2005). Van Stel et al. (2005), like Wennekers et al. (2005), are interested in the effect of entrepreneurship on national economic growth, while Audretsch and Thurik (2001) compare the new emerging entrepreneurial economy to the managed economy.

It is important to note that some of the most cited works are not research articles: the GEM global reports from the years 2001, 2004 and 2007, prepared by Reynolds et al. (2002), Acs et al. (2005), and Bosma et al. (2008), respectively. The first of these reports is by far the most cited GEM report. These works each carry out a cross-sectional study comparing entrepreneurship in various countries. Specifically, they study the rate of economic activity and its impact on economic growth, the characteristics of the new firms, and of the entrepreneurial population, and the profile and entrepreneurial potential of the rest of the population, and offer experts' opinions about the contextual factors of the environment that may have positive or negative repercussions for new firm creation.

Looking at the journals that published the cited works, by far the most important is Small Business Review, which published more than 53% of the studies making up the intellectual base of the GEM research. In particular, volume 24(3) concentrates two thirds of that journal's GEM-related studies.

It is also interesting to note that although co-authors do not appear in the graph (only the first author appears) various co-authors contribute to a large number of the studies discussed above, in particular Thurik, A.R., Wennekers, S., Reynolds, P., and Autio, E.

With regard to the themes dealt with in the intellectual base of the GEM project according to the graph, the main theme is the influence of entrepreneurship on national economic growth (Wennekers and Thurik, 1999; Wong et al., 2005; Carree et al., 2002; Wennekers et al., 2005; Van Stel et al., 2005). Visually these works concentrate in the left half of the graph, with few exceptions.

Fewer works focus on describing the GEM methodology (Reynolds et al., 2005; Sternberg and Wennekers, 2005) and the impact on entrepreneurship of variables such as the capacity to detect opportunities (Shane and Venkataraman, 2000) and other perceptual variables (Arenius and Minniti, 2005), and social and human capital (Davidsson and Honig, 2003). These works are mainly in the right half of the graph.

**5.2 Intellectual structure of PSED corpus**

Figure 6 shows the intellectual base of the PSED project, with only those pairs of works co-cited at least four times by the citing sample being retained. The co-citation threshold is lower for PSED than for GEM because its citing sample is smaller and because the authors found that with this co-citation threshold the resulting graph is visually more interpretable.

Figure 6. Structure of intellectual base of SSCI production in PSED project (raw co-citation≥4)

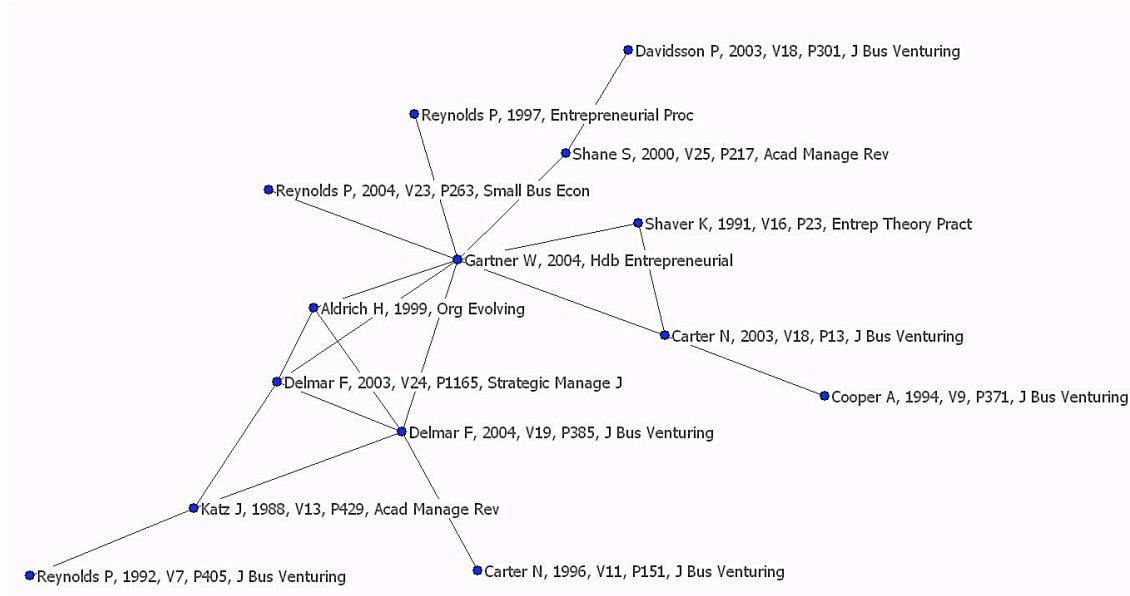

Analyzing the intellectual base of the PSED work, Gartner et al. (2004) is the most important work, both because of its centrality due to the high number of co-citation links shown in the graph, and because it is by far the most cited work in the research founded on PSED. Strikingly, this work is a book and not an article published in an academic journal. Nevertheless, the importance of this book, measured by its academic influence – which the graph shows – lends it sufficient legitimacy to belong to the intellectual base of the PSED research. This monographic work describes the fundamental principles of how the PSED panel works and studies the firm creation process and the impact of the individuals' life context, personal background and cognitive characteristics and the characteristics of the entrepreneurial environment on the firm's success or failure.

After the inevitable Gartner et al. (2004), PSED researchers mainly consider, and in this order: Carter et al. (2003), who analyze the reasons for becoming an entrepreneur (self-realization, financial success, roles, innovation, recognition, and independence); Delmar and Scott (2003), who stress the importance of drawing up a formal business plan; Shane and Venkataraman (2000), who analyze the existence, discovery and decision to exploit business opportunities; Reynolds and White (1997), who propose a model of the entrepreneurial process in four stages (conception, gestation, infancy, and adolescence) that considers the entrepreneur's psychological and social factors, contextual factors of the environment, and the impact of new firms on economic growth; Delmar and Scott (2004), who show that activities that generate legitimacy reduce the risk of failure; Aldrich (1999), who develops the evolutionary theory of organizations; Reynolds et al. (2004), who describe the PSED panel and analyze demographic differences between the nascent entrepreneurs; and Shaver and Scott (1991), who analyze the decision to create a business from a psychological perspective.

Other studies that form part of the intellectual base of PSED are: Cooper et al. (1994), who develop a predictive model of the nascent firm's probability of failure, survival and growth based on measures of its financial and human capital; Katz and Gartner (1988), who explore certain characteristics in the development of emerging organizations (intentionality, combination of resources, establishment of boundaries, and generation of exchanges); Carter et al. (1996), who analyze the activities of the nascent entrepreneur and whether these activities succeeded in setting the firm up in operation, or whether the entrepreneur had given up or was still trying to establish a firm; and Reynolds and Miller (1992), who study the gestation of the firm, and analyze its implications for research and the future development of the firm.

## 6. CONCLUSIONS

The GEM and PSED projects are two important milestones in the history of the research on entrepreneurship. Both show the importance of collaboration among a community of researchers for advancing knowledge about a phenomenon that influences economic development and the level of innovation in a society. It is important to deepen our understanding of the characteristics of these two projects to be able to manage and improve them. In this respect, the main contribution of the current work is its novel analysis of the scientific production generated by these two observatories.

Using bibliometric techniques on the published articles generated around these two projects and contained in the SSCI, the authors identified the most influential studies and authors, and analyzed their relations and the characteristics of the publications in each project. The authors analyzed the works most cited by the articles found in the SSCI relating to the two projects.

GEM investigates the relation between entrepreneurship and the economic growth of countries. It considers the cultural, political and social factors that influence the entrepreneurial environment, the specific conditions that affect firm creation and entrepreneurial potential, with this being understood in terms of motivation and perception of business opportunities. Its sources of information are surveys of the population, interviews with experts and secondary sources.

The PSED project obtains its primary data from a panel of individuals involved in firm creation. Unlike GEM, which is a cross-sectional study, PSED is a longitudinal study in which individuals remain over the years and are interviewed at regular intervals. The aim is to obtain information about the characteristics of the process, the population involved in new firm creation, and the activities and attributes of the nascent firms. Its intellectual base uses demographic and cognitive approaches. Furthermore, it uses research into opportunity recognition, the networks and resources forming part of the entrepreneurial process, and the entrepreneurial environment.

The two projects partially coincide in their approaches, and hence the publications based on them show some similarities. In both groups, the psycho-sociological and demographic aspects of the entrepreneur are important, as are the contextual factors of the environment.

But significant differences exist between the two groups of work. Although both are interested in the entrepreneurial process, the PSED studies take a broader look at this process, being more interested than their GEM counterparts in the stages prior to the founding of the firm and how the entrepreneurial process evolves over time. This latter point is explained by the fact that PSED's methodology is based on a longitudinal study. The PSED studies are also more interested in the entrepreneurial process and on whether this influences the growth, establishment or failure of the new firm. This is not to say that GEM studies ignore these questions, but rather that the PSED research is more interested in looking into the black box of the entrepreneurial process, the characteristics of the people, the resources, and the stages of the process.

In turn, GEM's intellectual base focuses more on the impact of firm creation on economic growth in countries and regions. This is explained by the fact that GEM's empirical study is cross-sectional and that the greater the number of participating countries the better.

Looking at the two projects at greater depth, these differences remain. Considering just the co-citation thresholds used in the graphs in this work, the intellectual bases of the two projects coincide in only two studies. The first of these works is Davidsson and Honig (2003), which analyses social and human capital as predictors of entrepreneurial behavior. The second is Shane and Venkataraman (2000), which lays the foundations of the discipline of entrepreneurship and so occupies a central position in the intellectual bases of both PSED and GEM.

It is also noteworthy that both intellectual bases frequently cite works in which Professor Paul Reynolds participates. Oddly though, none of his works is present in both intellectual bases simultaneously (at least with the citation thresholds chosen here). Nevertheless, this prestigious researcher has clearly had a profound influence in the construction and development of the intellectual bases of both projects.

The differences in the intellectual bases of the two projects are also evident in how close their studies are to publications with more or less traditional approaches in the field of entrepreneurship. According to Fisher, the traditional model is mainly founded on the economic approach, on competitive analysis, on prediction as a method to reduce uncertainty and exercise control, on rational decision-making, and on establishing a plan for creating a firm. Other emerging perspectives, such as the effectuation approach or bricolage suggest that in some contingencies the entrepreneurs change their routes to identify, create and exploit opportunities, with action and co-creation characterizing the process (Fisher, 2012).

The publications of each project analyzed here show some signs of these two approaches. In the GEM group, studies framed in a more traditional approach are more frequent. Thus the focus is on topics like economic growth (Wennekers, and Thurik, 1999; Wennekers et al., 2005; Wong et al., 2005; Van Stel et al., 2005), prediction or the determinants conditioning entrepreneurship (Davidsson and Honig, 2003; Arenius and Minniti, 2005; Sternberg and Wennekers, 2005) or opportunity discovery and exploitation (Shane and Venkataraman, 2000), this latter being one of the most cited works in both projects. This perspective is also present in other publications from the PSED group. Thus some works stress the importance of the rational and predictive side (Delmar and Scott, 2003; Cooper et al., 1994; Davidsson and Honig, 2003) or the determinants of entrepreneurship (Reynolds and Miller, 1992; Reynolds et al., 2004b).

The publications related to emergent approaches are less frequent in both projects, but the PSED group is closer to these approaches. Only one publication in the GEM group can be labeled this way: Audretsch and Thurik (2001). These authors identify the characteristics of the new emerging economy, which they call the entrepreneurial economy, compared to what they call the managed economy. PSED's core publication (Gartner et al., 2004) hints at emerging theoretical concepts close to social and co-creation aspects. Other PSED publications also use concepts from the emergent literature such as the use of resources, their combination and exchange (Katz and Gartner, 1988), or the dynamic approach of organizational knowledge and social change (Aldrich, 1999).

Nevertheless, the majority of publications related to emergent theories are recent and consequently have had less chance of being used by either project, so these results must be regarded as exploratory and open to modification in the future as new publications appear and new theoretical frameworks are consolidated.

The main limitations of this work are a consequence of the research design and the bibliometric techniques used. Among the first is the criterion for selecting the corpus of work to represent the scientific production generated by each observatory, based on the search for keywords in a single database (SSCI). The conclusions reached must have a limited generalizability because the documents analyzed are only a small part of the published results in each observatory. Nevertheless, the authors are reasonably convinced that the literature analyzed is representative of the most important research in this discipline.

The use of bibliometric techniques inevitably has its limitations. Thus for example, when counting citations it is impossible to consider the reasons why authors cite works. They may do this for various reasons: when looking at previous work and building a theoretical framework, to criticize a reference, to demonstrate their knowledge, to give the impression they have used more literature, or to cite their own work. On the other hand, works are not cited again for various reasons: because they have achieved such acceptance by the scientific community that authors no longer feel any need to cite the work (obliteration), or for other, less noble reasons. The authors partially get round these limitations by the strict review process of the journals analyzed.

The citations were also counted during a particular period of time. Thus the studies published at the end of this period were available to the scientific community for a shorter time and so had less chance of being cited. This phenomenon is indisputable, but since the authors have considered that the number of citations obtained is not an indicator of quality but rather of impact or influence, they believe that it is logical that this number reflects the fact that more recent studies have not had enough time to have an impact on the literature of the discipline.

Co-citation analysis also has its limitations. This method allows researchers to represent only a part of the cited literature, although this part is considered the most representative, and interpretation of the graphs obtained is necessarily subjective. In any case, although some works are absent the clusters of documents are testimony to the existence of a group of researchers who share the same interests and the same references (Callon et al., 1993).

Some of these limitations have no solution, but are not exclusive to this type of tool, since they appear in other non-experimental disciplines. Others can be resolved, and future researchers would be encouraged to improve the techniques applied here. In this respect, the current authors aim to expand the citing sample to include other citation indexes (e.g., SCOPUS), increase the number of documents used to identify the intellectual base by raising the citation thresholds, and improve their interpretation by exploiting the explanatory power of methods from social network analysis to identify groups of homogeneous documents and measure their density and centrality in the network of co-citations in which they are immersed.

Finally, exploring the proximity of the studies of both projects to more traditional or emerging approaches would open up new lines of research relating to conceptual aspects and schools of thought.

Wong, P.K., Yuen, P. H., and Autio, E. (2005). Entrepreneurship, innovation and economic growth: Evidence from GEM data. Small Business Economics, 24(3), 335-350.

World Bank (2002), World Development Indicators 2002, International Bank for Reconstruction and Development, Washington, DC.

World Economic Forum (2002), Global Competitiveness Report 2001-2002, Oxford, Oxford University Press.

Ying, D., Gobinda, C., Shubert, F. (1999). Mapping the Intellectual Structure of Information Retrieval Studies: an Author Co-citation Analysis 1987-1997. *Journal of Information Science*, 25, 67-78.31